\documentclass[11pt,twoside]{article}
\usepackage{./asp2014}

\aspSuppressVolSlug
\resetcounters

\begin{document}

\title{The Benchmark Eclipsing Binary V530 Ori: A Critical Test of
  Magnetic Evolution Models for Low-Mass Stars}

\author{
Guillermo Torres$^1$,
Claud H.\ Sandberg Lacy$^2$,
Kre\v{s}imir Pavlovski$^3$,
Gregory A.\ Feiden$^4$,
Jeffrey A.\ Sabby$^{5,6}$,
Hans Bruntt$^{7,8}$, and
Jens Viggo Clausen$^9$
\affil{$^1$Harvard-Smithsonian Center for Astrophysics, Cambridge, MA
  02138, USA; \email{gtorres@cfa.harvard.edu}}
\affil{$^2$Department of Physics, University of Arkansas,
  Fayetteville, AR 72701, USA}
\affil{$^3$Department of Physics, Faculty of Science, University
  of Zagreb, 10000 Zagreb, Croatia}
\affil{$^4$Department of Physics \& Astronomy, Uppsala
  University, SE-751 20 Uppsala, Sweden}
\affil{$^5$Physics Department, Southern Illinois University
  Edwardsville, Edwardsville, IL 62026, USA}
\affil{$^6$Visiting Astronomer, Kitt Peak National Observatory,
  National Optical Astronomy Observatories, operated by the
  Association of Universities for Research in Astronomy, Inc., under a
  cooperative agreement with the National Science Foundation.}
\affil{$^7$Stellar Astrophysics Centre, Department of Physics and
  Astronomy, Aarhus University, 8000 Aarhus C,
  Denmark}
\affil{$^8$Aarhus Katedralskole, 8000 Aarhus C, Denmark}
\affil{$^9$Niels Bohr Institute, Copenhagen University, DK-2100
  Copenhagen {\O}, Denmark --- deceased}}

\paperauthor{Torres, G.}{gtorres@cfa.harvard.edu}{}{Harvard-Smithsonian Center for Astrophysics}{}{Cambridge}{Massachusetts}{02138}{USA}
\paperauthor{Claud H.\ Sandberg Lacy}{}{}{University of Arkansas}{Department of Physics}{Fayetteville}{AR}{AR 72701 }{USA}
\paperauthor{Kre\v{s}imir Pavlovski}{}{}{University
  of Zagreb}{Department of Physics, Faculty of Science}{Zagreb}{}{10000}{Croatia}
\paperauthor{Gregory~A.~Feiden}{gregory.a.feiden@gmail.com}{0000-0002-2012-7215}
{Uppsala University}{Department of Physics \& Astronomy}{Uppsala}{}{SE-751 20}{Sweden}
\paperauthor{Jeffrey A.\ Sabby}{}{}{Southern Illinois University}{Physics Department}{Edwardsville}{IL}{IL 62026}{USA}
\paperauthor{Hans Bruntt}{}{}{Aarhus University}{Stellar Astrophysics Centre, Department of Physics and
  Astronomy}{Aarhus}{}{8000}{Denmark}
\paperauthor{Jens Viggo Clausen}{}{}{Copenhagen University}{Niels Bohr Institute}{Copenhagen}{}{}{Denmark}

\begin{abstract}
We report accurate measurements of the physical properties (mass,
radius, temperature) of components of the G+M eclipsing binary
V530\,Ori. The M-type secondary shows a larger radius and a cooler
temperature than predicted by standard stellar evolution models, as
has been found for many other low-mass stars and ascribed to the
effects of magnetic activity and/or spots. We show that models from
the Dartmouth series that incorporate magnetic fields are able to
match the observations with plausible field strengths of 1-2 kG,
consistent with a rough estimate we derive for that star.
\end{abstract}

%%%%%%%%%%%%%%%%%%%%%%%%%%%%%%%%%%%%%%%%%%%%%%%%%%%%%%%%%%%%%%%%%%%%%%
\section{Introduction}
\label{sec:introduction}
%%%%%%%%%%%%%%%%%%%%%%%%%%%%%%%%%%%%%%%%%%%%%%%%%%%%%%%%%%%%%%%%%%%%%%

It has been known since the 1970's
% \citep{Hoxie:70, Hoxie:73, Lacy:77}
that many low-mass stars in eclipsing binaries tend to exhibit larger
radii and cooler temperatures than predicted by standard models of
stellar evolution \citep[for a recent review, see][]{Torres:13}
(or Feiden's review in this volume). This
phenomenon is widely believed to be the result of stellar activity
(magnetic fields and/or spots). Efforts to understand the anomalies by
incorporating magnetic fields into the stellar evolution models have
made good progress in the last decade or so \citep[e.g.,][]{Mullan:01,
  Chabrier:07, Feiden:12}, indicating such models are capable of
producing ``radius inflation'' and ``temperature suppression'' of
approximately the right magnitude to explain the observed deviations.
However, more quantitative comparisons are generally difficult because
the ages and chemical composition of the low-mass stars in eclipsing
binaries with well measured masses, radii, and effective temperatures
are typically unknown, as are the magnetic field strengths of the
components.

Here we present an analysis of the eclipsing binary system V530\,Ori,
composed of a G-type primary star and an M-type secondary in a
6.11-day orbit. This combination presents a number of advantages: the
solar-type primary facilitates the measurement of the metallicity and
other characteristics of the system, and while faint (only 1--2\% of
the flux of the primary), the low-mass secondary is still detectable
spectroscopically and allows an accurate measurement of its
fundamental properties.  Most other well-studied eclipsing binaries
with low-mass stars contain two M dwarfs, which makes it challenging
to establish the metallicity given the complicated nature of M-star
spectra.

%%%%%%%%%%%%%%%%%%%%%%%%%%%%%%%%%%%%%%%%%%%%%%%%%%%%%%%%%%%%%%%%%%%%%%
\section{Observations}
\label{sec:observations}
%%%%%%%%%%%%%%%%%%%%%%%%%%%%%%%%%%%%%%%%%%%%%%%%%%%%%%%%%%%%%%%%%%%%%%

Extensive spectroscopy of V530\,Ori was obtained using three different
telescope and instrument setups at the Harvard-Smithsonian Center for
Astrophysics and at the Kitt Peak National Observatory from 1996 to
2014, totaling 145 high-resolution spectra.  Radial velocities for
both components were measured using the two-dimensional
cross-correlation technique TODCOR \citep{Zucker:94}, in the same
manner as described recently by \cite{Sandberg:14}.

More than 8000 differential $V$-band photometric observations of
V530\,Ori were collected with two robotic telescopes (URSA and NFO)
from 2001 to 2012, operating at the University of Arkansas and in New
Mexico (USA), respectively.  The telescopes, instrumentation, data
acquisition and reduction have been described by \cite{Grauer:08} and
\cite{Sandberg:12}. We additionally obtained 720 $uvby$ measurements
with the Str\"omgren Automatic Telescope at ESO (La Silla, Chile) from
2001 to 2006. A description of the reduction procedures for these
observations is given by \cite{Clausen:08}.

%%%%%%%%%%%%%%%%%%%%%%%%%%%%%%%%%%%%%%%%%%%%%%%%%%%%%%%%%%%%%%%%%%%%%%
\section{Analysis and Results}
\label{sec:analysis}
%%%%%%%%%%%%%%%%%%%%%%%%%%%%%%%%%%%%%%%%%%%%%%%%%%%%%%%%%%%%%%%%%%%%%%

The three radial-velocity data sets were combined into a single
solution for the spectroscopic elements, after verifying that
individual solutions are consistent with each other. The fit
indicates the orbit is slightly eccentric ($e = 0.08802 \pm 0.00023$),
and gives a mass ratio of $q \equiv M_{\rm B}/M_{\rm A} = 0.5932 \pm
0.0022$.  Light-curve solutions were performed with the JKTEBOP code
\citep{Nelson:72, Popper:81, Southworth:04}, and indicate the
secondary eclipse is total (totality duration of 70 minutes), and the
primary is annular. Our spectroscopic orbital solution and sample fits
to the URSA and NFO photometry are shown in Figure~\ref{fig:obs}.

% Fig.1
\articlefigurethree{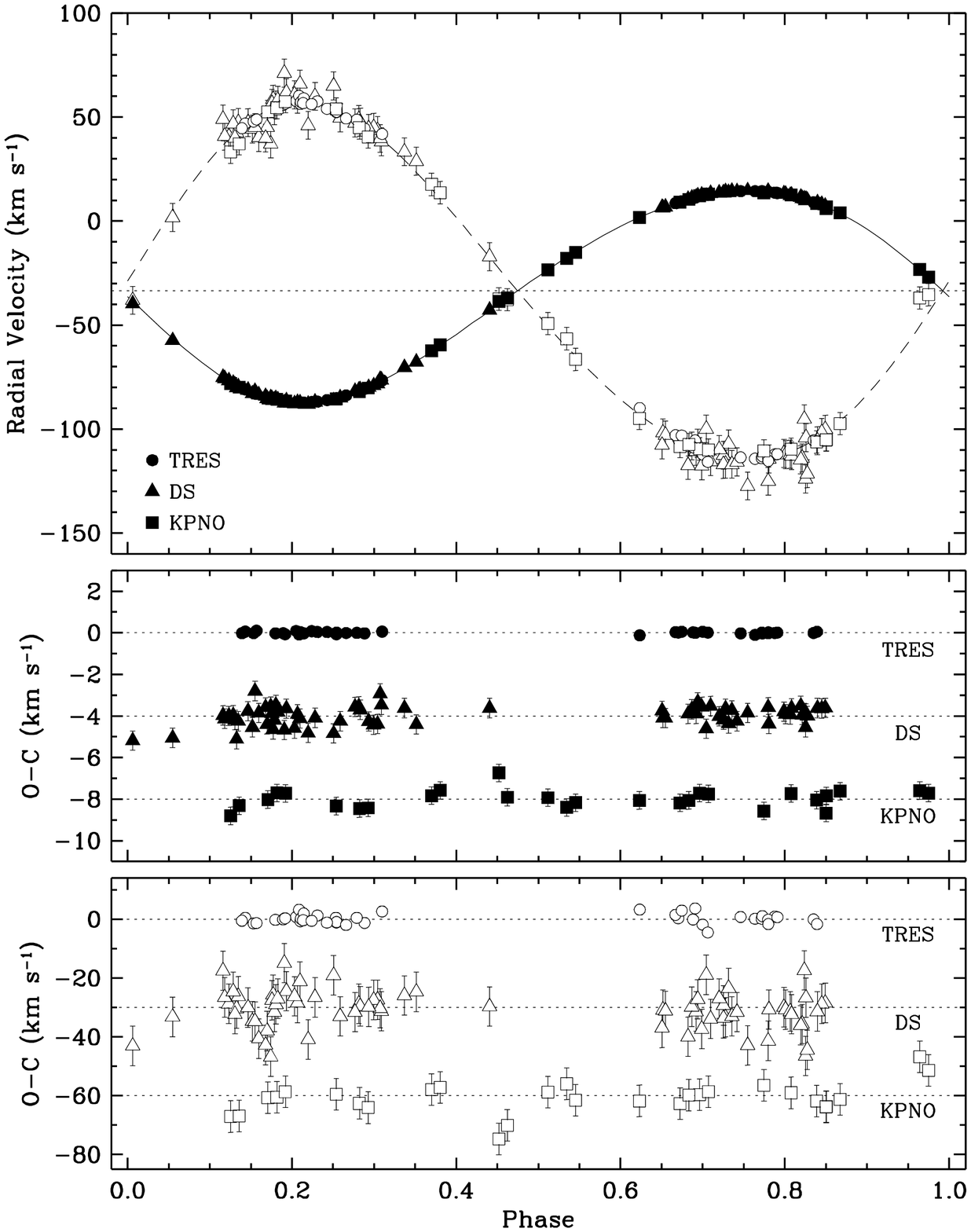}{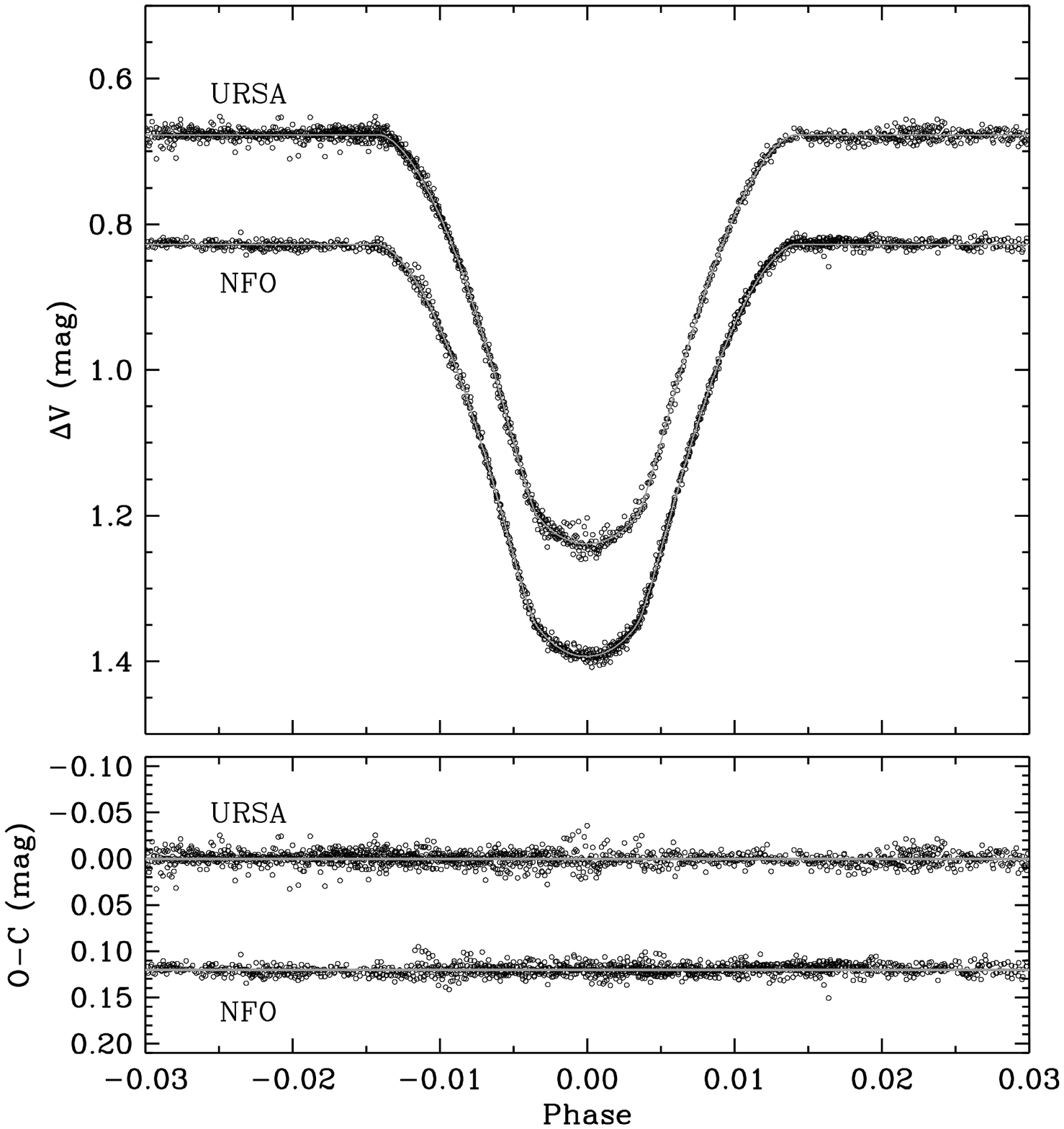}{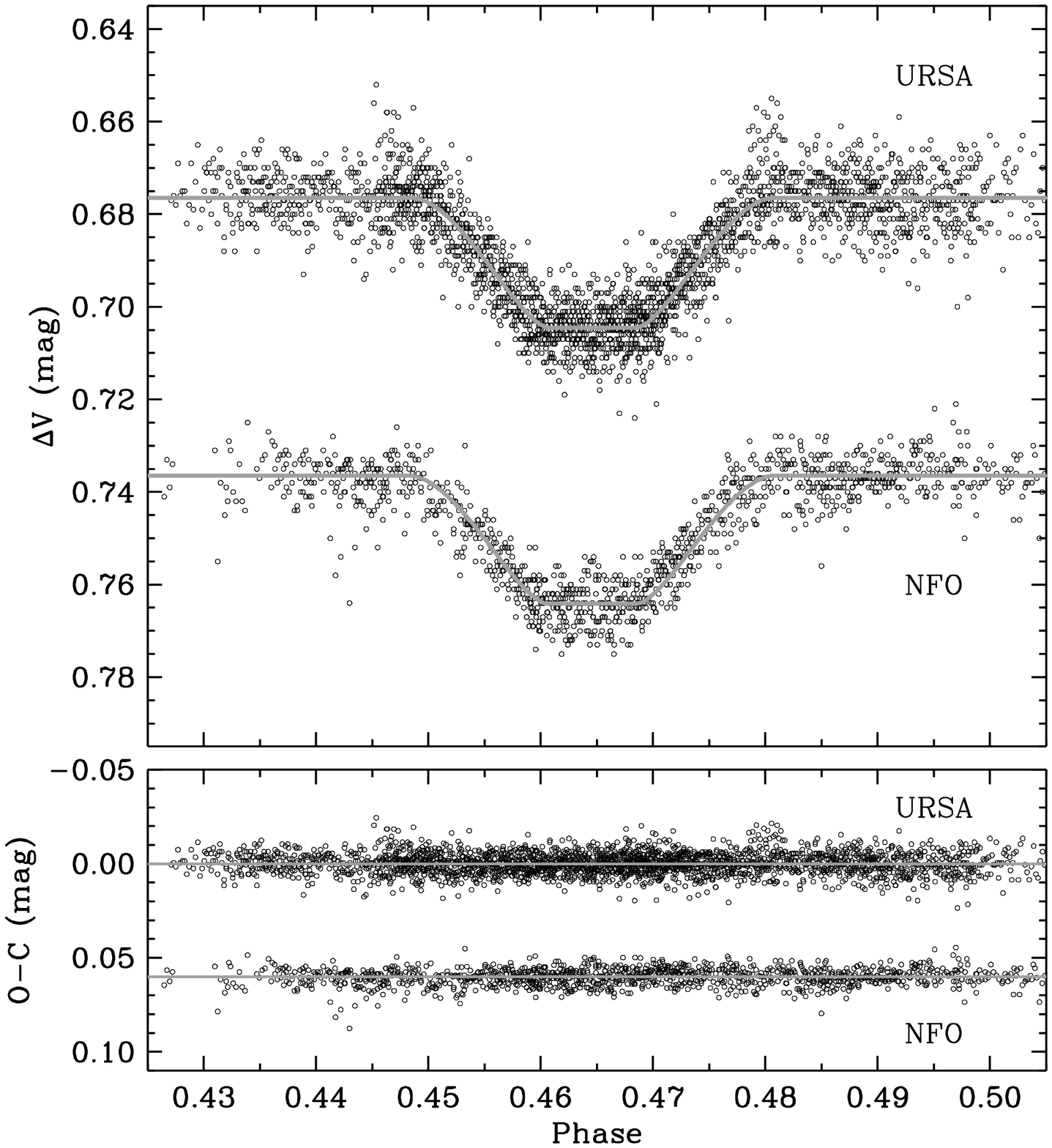}
{fig:obs}{\emph{Left:}
  Measured radial velocities and combined spectroscopic orbital
  solution, with residuals shown at the bottom; \emph{Middle and
    right:} URSA and NFO photometric observations near the primary
  (middle) and secondary (right) eclipses, along with our best fit
  model. Residuals are displayed at the bottom.}

We also subjected the three sets of spectra to disentangling using the
{\sc FDBinary} program \citep{Ilijic:04}. The disentangled primary
spectrum was then analyzed with the {\sc uclsyn} package
\citep{Smalley:11} to obtain estimates of the primary temperature as
well as chemical abundances of iron, ${\rm [Fe/H]} = -0.12 \pm 0.08$,
and 20 other elements. Several other spectroscopic estimates of the
primary temperature are in good agreement. The secondary temperature was
derived on the basis of the primary value and the central surface
brightness ratio inferred from the light curve analysis. The final
absolute dimensions of the system are presented in
Table~\ref{tab:absdim}, and show that the masses and radii are good to
about 1\%.  The primary star is very similar in mass to the Sun.

\begin{table}
\caption{Physical properties of the components of V530\,Ori.\label{tab:absdim}}
\smallskip
\begin{center}
{\small
\begin{tabular}{lcc}
\tableline
\noalign{\smallskip}
~~~~~~~~Parameter~~~~~~~~ & Star A & Star B \\
\noalign{\smallskip}
\tableline
\noalign{\smallskip}
Mass ($M_{\sun}$)\dotfill     & 1.0038~$\pm$~0.0066      & 0.5955~$\pm$~0.0022   \\
Radius ($R_{\sun}$)\dotfill   &  0.980~$\pm$~0.013       & 0.5873~$\pm$~0.0067   \\
$\log g$ (cgs)\dotfill        &  4.457~$\pm$~0.012       &  4.676~$\pm$~0.010    \\
$T_{\rm eff}$ (K)\dotfill     &   5890~$\pm$~100\phantom{2}     &   3880~$\pm$~120\phantom{2}  \\
$\log L/L_{\sun}$\dotfill       &  0.016~$\pm$~0.032       & $-$1.154~$\pm$~0.053\phantom{$-$}    \\
$M_V$ (mag)\dotfill &   4.71~$\pm$~0.10        &   8.72~$\pm$~0.11     \\
$E(B-V)$ (mag)\dotfill        &      \multicolumn{2}{c}{0.045~$\pm$~0.020}       \\
Distance (pc)\dotfill        &      \multicolumn{2}{c}{103~$\pm$~6\phantom{22}}    \\
${\rm [Fe/H]}$\dotfill                        & \multicolumn{2}{c}{$-0.12 \pm 0.08$\phantom{$-$}} \\
\noalign{\smallskip}
\tableline
\end{tabular}
}

\end{center}
\end{table}

\vspace{-6mm}
%%%%%%%%%%%%%%%%%%%%%%%%%%%%%%%%%%%%%%%%%%%%%%%%%%%%%%%%%%%%%%%%%%%%%%
\section{Stellar Evolution Models and Magnetic Fields}
\label{sec:models}
%%%%%%%%%%%%%%%%%%%%%%%%%%%%%%%%%%%%%%%%%%%%%%%%%%%%%%%%%%%%%%%%%%%%%%

\begin{figure}
\centering
\begin{tabular}{cc}
\includegraphics[width=6.0cm]{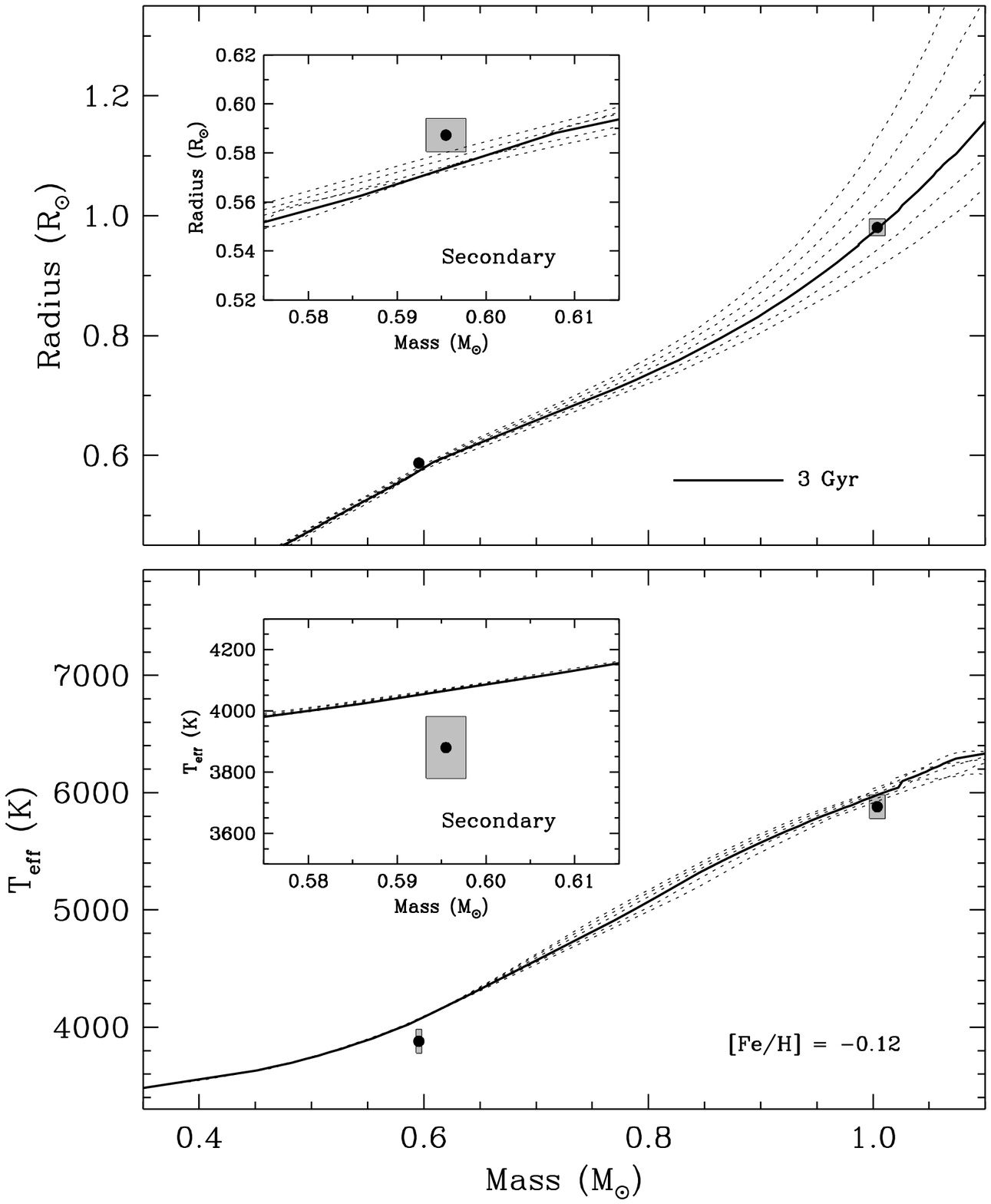} &
\includegraphics[trim = 0pt 0pt 190pt 0pt, clip, width=6.1cm]
{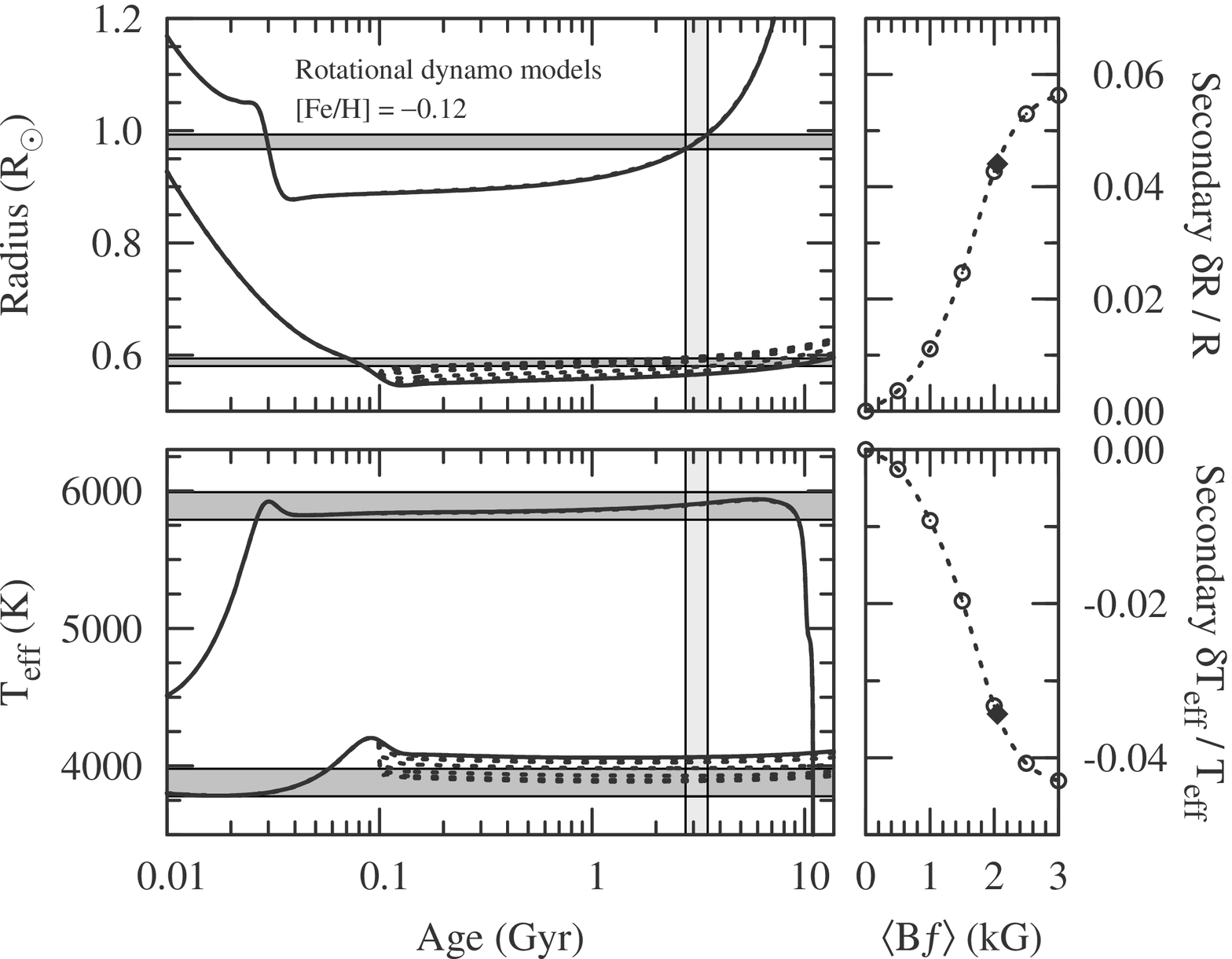} \\
\end{tabular}

\caption[]{\emph{Left:} Observations for V530\,Ori compared against
  1--6\,Gyr isochrones from the standard Dartmouth models
  \citep{Dotter:08} for the measured metallicity. The insets zoom in
  on the secondary. \emph{Right:} Horizontal bands show the radius and
  temperature measurements. Solid lines are standard evolutionary
  tracks from the Dartmouth models for the measured masses, and dotted
  lines incorporate magnetic fields for the secondary using the
  rotational dynamo prescription, with field strengths between 0.5 and
  3.0\,kG.\label{fig:models}}

\end{figure}

A comparison of the measured masses, radii, and temperatures of both
stars with standard (non-magnetic) isochrones from the Dartmouth
series \citep{Dotter:08} for the measured metallicity shows excellent
agreement for the primary at an age of about 3\,Gyr, but not for the
secondary, which appears larger than predicted by $\sim$3.7\% and
cooler by $\sim$4.8\% (see Figure~\ref{fig:models}, left).  Good
agreement for the primary is also seen with the Yonsei-Yale models
\citep{Yi:01, Demarque:04}.

An example of a comparison with Dartmouth models that incorporate
magnetic fields \citep{Feiden:13} is presented in the right panel of
the figure, using the rotational or shell dynamo prescription
($\alpha$--$\Omega$), in which the field is rooted in the strong shear
layer at the base of the convective zone (tachocline). We find that
with this formulation magnetic fields have relatively little effect on
the primary properties for modest magnetic field strengths of
$\sim$170\,G, such as we estimate for this star below, but do allow a
good fit to the secondary properties at the same age as the primary
for a field strength of $\langle Bf\rangle = 2.1 \pm 0.4$\,kG.

% Fig.2
\setlength{\tabcolsep}{5pt}
\setlength{\tabcolsep}{6pt}

An alternate prescription for the secondary using a turbulent or
distributed dynamo ($\alpha^2$) also provides a good match to its
properties with a field strength of $\langle Bf\rangle = 1.3 \pm
0.4$\,kG, at the same age as the primary. In this case the energy for
the magnetic field is drawn from the large-scale convective flow.
However, this type of dynamo has a larger effect on the primary
properties, such that a 170\,G field strength for that star produces a
poorer fit to its properties at a somewhat younger age of 2.4\,Gyr,
and although still possible, reproducing the secondary properties at
the same age becomes more difficult.

Are these predicted magnetic field strengths for the secondary
plausible? To answer this question, we estimated the field strengths
for both stars using an empirical relation between $\langle Bf\rangle$
and the inverse Rossby number proposed by \cite{Saar:01}. We assumed
the stars are rotationally synchronized, which is reasonable given
that the timescale for this phenomenon is much shorter than the age
for this system. With Rossby numbers of 0.431 and 0.116 for the
primary and secondary based on convective turnover times from the same
source used by \cite{Saar:01}, we obtained rough $\langle Bf\rangle$
values of $170 \pm 140$\,G and $830 \pm 650$\,G. The latter estimate
is indeed consistent with theoretical predictions, within the
admittedly large uncertainties.

%%%%%%%%%%%%%%%%%%%%%%%%%%%%%%%%%%%%%%%%%%%%%%%%%%%%%%%%%%%%%%%%%%%%%%
\section{Conclusions}
\label{sec:conclusions}
%%%%%%%%%%%%%%%%%%%%%%%%%%%%%%%%%%%%%%%%%%%%%%%%%%%%%%%%%%%%%%%%%%%%%%

V530\,Ori is an important new benchmark eclipsing binary system with
accurately measured properties containing a solar-type primary and
low-mass secondary displaying radius inflation and temperature
suppression, when compared to standard evolution models. Magnetic
models are able to match the secondary properties with plausible
magnetic field strengths of 1--2\,kG, suggesting we are on the right
track to understanding these discrepancies. Based strictly on the
quality of the fits, the observations seem to suggest a scenario in
which magnetic fields have only a minor effect on the solar-mass
primary, consistent with it having a rotational dynamo, which is also
believed to be mechanism operating in the Sun. The nature of the
magnetic field on the secondary is less clear, with the observations
perhaps favoring a turbulent ($\alpha^2$) dynamo over a rotational
one, but not at a very significant level.

\acknowledgements We thank the observers on Mount Hopkins and La Silla
for their assistance, as well as Bill Neely for the operation and
maintenance of the NFO. JVC participated fully in the data collection
and analysis up to the time of his death on 2011 June 5, but bears no
responsibility for the final text of this paper. GT acknowledges
partial support for this work from NSF grant AST-1007992.

%%%%%%%%%%%%%%%%%%%%%%%%%%%%%%%%%%%%%%%%%%%%%%%%%%%%%%%%%%%%%%%%%%%%%%


\begin{thebibliography}
%%%%%%%%%%%%%%%%%%%%%%%%%%%%%%%%%%%%%%%%%%%%%%%%%%%%%%%%%%%%%%%%%%%%%%

\bibitem[Chabrier et al.(2007)]{Chabrier:07} Chabrier, G., Gallardo,
  J., \& Baraffe, I. 2007, \aap, 472, L17

\bibitem[Clausen et al.(2008)]{Clausen:08} Clausen, J.\ V., Vaz,
  L.\ P.\ R., Garc\'\i a, J.\ M., et al. %Gim\'enez, A., Helt, B.\ E., Olsen,
  %E.\ H., \& Andersen, J. 
  2008, \aap, 487, 1081

\bibitem[Demarque et al.(2004)]{Demarque:04} Demarque, P., Woo, J.-H.,
  Kim, Y.-C., \& Yi, S.\ K.\ 2004, \apjs, 155, 667

\bibitem[Dotter et al.(2008)]{Dotter:08} Dotter, A., Chaboyer, B.,
  Jevremovi{\'c}, D., et al.\ 2008, \apjs, 178, 89

\bibitem[Feiden \& Chaboyer(2012)]{Feiden:12} Feiden, G.\ A., \&
  Chaboyer, B. 2012, \apj, 761, 30

\bibitem[Feiden \& Chaboyer(2013)]{Feiden:13} Feiden, G.\ A., \&
  Chaboyer, B. 2013, \apj, 779, 183

\bibitem[F\H{u}r\'esz(2008)]{Furesz:08} F\H{u}r\'esz, G. 2008, PhD
  thesis, Univ. Szeged, Hungary

\bibitem[Grauer et al.(2008)]{Grauer:08} Grauer, A.\ D., Neely,
  A.\ W., \& Lacy, C.\ H.\ S. 2008, \pasp, 120, 992

\bibitem[Iliji\'{c} et al.(2004)]{Ilijic:04} Iliji\'{c}, S.,
  Hensberge, H., Pavlovski, K., \& Freyhammer, L.\ S. 2004,
  Spectroscopically and Spatially Resolving the Components of the
  Close Binary Stars, eds.\ R.\ W.\ Hilditch, H.\ Hensberge and
  K.\ Pavlovski (San Francisco:ASP), ASP Conf.\ Ser., 318, 111

\bibitem[Latham(1992)]{Latham:92} Latham, D.\ W. 1992, in IAU
  Coll.\ 135, Complementary Approaches to Double and Multiple Star
  Research, ASP Conf.\ Ser.\ 32, eds.\ H.\ A.\ McAlister \&
  W.\ I.\ Hartkopf (San Francisco: ASP), 110

\bibitem[Mullan \& MacDonald(2001)]{Mullan:01} Mullan, D.\ J.,
  MacDonald, J. 2001, \apj, 559, 353

\bibitem[Nelson \& Davis(1972)]{Nelson:72} Nelson, B., \& Davis,
  W.\ D. 1972, \apj, 174, 617

\bibitem[Popper \& Etzel(1981)]{Popper:81} Popper, D.\ M., \& Etzel,
  P.\ B. 1981, \aj, 86, 102

\bibitem[Saar(2001)]{Saar:01} Saar, S.\ H. 2001, 11th Cambridge
  Workshop on Cool Stars, Stellar Systems and the Sun,
  eds.\ R.\ J.\ Garc\'\i a L\'opez, R.\ Rebolo and M.\ R.\ Zapatero
  Osorio (San Francisco: ASP), ASP Conv.\ Ser., 223, 292

\bibitem[Sandberg Lacy et al.(2012)]{Sandberg:12} Sandberg Lacy, C.\ H.,
  Torres, G., \& Claret, A. 2012, \aj, 144, 167

\bibitem[Sandberg Lacy et al.(2014)]{Sandberg:14} Sandberg Lacy,
C.\ H., Torres, G., Fekel, F.\ C., \& Muterspaugh, M.\ W. 2014, \aj,
147, 148

\bibitem[Smalley et al.(2011)]{Smalley:11} Smalley, B., Smith, K.\ C.,
  \& Dworetsky, M.\ M. 2001, {\sc uclsyn} User Guide, available at
     {\footnotesize \url{http://www.astro.keele.ac.uk/~bs/publs/uclsyn.pdf}}

\bibitem[Southworth et al.(2004)]{Southworth:04} Southworth, J.,
  Maxted, P.\ F.\ L., \& Smalley, B. 2004, \mnras, 349, 547

\bibitem[Torres(2013)]{Torres:13} Torres, G. 2013, AN, 334, 4

\bibitem[Yi et al.(2001)]{Yi:01} Yi, S., Demarque, P., Kin, Y.-C., et al. 
  %Lee, Y.-W., Ree, C. H., Lejeune, T., \& Barnes, S. 
  2001, \apjs, 136, 417

\bibitem[Zucker \& Mazeh(1994)]{Zucker:94} Zucker, S., \& Mazeh, T.
  1994, \apj, 420, 806

\end{thebibliography}
\end{document}